\documentclass[conference]{IEEEtran}
\ifCLASSINFOpdf
\else
\fi
\usepackage[utf8]{inputenc}
\usepackage{amsmath}
\usepackage{amsthm}
\usepackage{cite}
\usepackage{mathtools}
\usepackage{amsfonts}
\usepackage{amssymb}
\usepackage{graphicx}
\usepackage{subfigure}
\usepackage{multirow}
\usepackage{float}
\usepackage{mathrsfs}
\usepackage{algorithm}
\usepackage[noend]{algpseudocode}
\usepackage{graphicx}
\usepackage{xcolor}
\usepackage{colortbl}
\usepackage{pbox}
\usepackage{lipsum}
\allowdisplaybreaks
\usepackage{algorithm}
\usepackage{algpseudocode}
\usepackage{algpseudocode}
\algnewcommand{\LineComment}[1]{\State \(\triangleright\) #1}
\usepackage[T1]{fontenc}
\usepackage[utf8]{inputenc}

\IEEEoverridecommandlockouts

\usepackage{verbatim}

\newtheorem{theorem}{\bf Theorem}

\def\href#1#2{#1 #2}

\ifCLASSINFOpdf

\else

\fi

\begin{document}
\bstctlcite{IEEEexample:BSTcontrol}
\title{\Huge Dependence Control for Reliability Optimization in Vehicular Networks\vspace{-0.8cm}}

\author{\IEEEauthorblockN{Tengchan Zeng\IEEEauthorrefmark{1}, Omid Semiari\IEEEauthorrefmark{2}, Walid Saad\IEEEauthorrefmark{1}, and Mehdi Bennis\IEEEauthorrefmark{3}}
	\IEEEauthorblockA{\IEEEauthorrefmark{1}Wireless@VT, Department of Electrical and Computer Engineering, Virginia Tech, Blacksburg, VA, USA}
	
	\IEEEauthorblockA{\IEEEauthorrefmark{2}Department of Electrical and Computer Engineering, University of Colorado Colorado Springs, Colorado Springs, CO, USA} 

	\IEEEauthorblockA{\IEEEauthorrefmark{3}Centre for Wireless Communications, University of Oulu, Oulu, Finland} 	
	Emails: \IEEEauthorrefmark{1}\{tengchan, walids\}@vt.edu, \IEEEauthorrefmark{2}osemiari@uccs.edu, 
	\IEEEauthorrefmark{3}mehdi.bennis@oulu.fi 
    \thanks{This research was supported, in part, by the U.S. National Science Foundation under Grants CNS-1739642, CNS-1836802 and CNS-1941348, by the Academy of Finland Project CARMA, by the Academy of Finland Project MISSION, by the Academy of Finland Project SMARTER, as well as by the INFOTECH Project NOOR.}
	\vspace{-0.5cm}
}

\maketitle

\begin{abstract}
	Vehicular networks will play an important role in enhancing road safety, improving transportation efficiency, and providing seamless Internet service for users on the road. 
	Reaping the benefit of vehicular networks is contingent upon meeting stringent wireless communication performance requirements, particularly in terms of delay and reliability.
	In this paper, a dependence control mechanism is proposed to improve the overall reliability of vehicular networks.
	In particular, the dependence between the communication delays of different vehicle-to-vehicle (V2V) links is first modeled. 
	Then, the concept of a \emph{concordance order}, stemming from stochastic ordering theory, is introduced to show that a higher dependence can lead to a better reliability. 
	Using this insight, a power allocation problem is formulated to maximize the concordance, thereby optimizing the overall communication reliability of the V2V system. 
	To obtain an efficient solution to the power allocation problem, a dual update method is introduced.
	Simulation results verify the effectiveness of performing dependence control for reliability optimization in a vehicular network, and show that the proposed mechanism can achieve up to $25$\% reliability gain compared to a baseline system that uses a random power allocation.   
\end{abstract}

\IEEEpeerreviewmaketitle

\section{Introduction}

Vehicular networks will be a pillar of tomorrow's intelligent transportation systems (ITSs) \cite{saad20196G}. 
In essence, vehicular networks allow vehicles to share traffic information and safety messages with each other, coordinate their mobility, and use high-speed Internet services.
By using reliable vehicular communication links, connected vehicles can obtain situational awareness of their surrounding environment which allows them to make safer decisions, thus increasing the overall road safety. 
Moreover, by sharing information among different vehicles, traffic congestion can be alleviated via intelligent traffic control and management.
However, to enable effective vehicular communication networks, many technical challenges must be addressed ranging from channel modeling \cite{6823640} and edge computing \cite{8647367} to privacy and security issues \cite{5307471}.

One of the most important challenges of vehicular communications is meeting stringent service requirements in terms of communication latency and reliability \cite{7572192,7990497,5GPPPautomotive}.
For example, when the long-term evolution (LTE) system is used to transmit hazard warning signals among vehicles, the transmission delay must be maintained below $100$~ms and the packet reception probability should exceed 95\% \cite{7572192}.
Moreover, for pre-crash sensing messages, the transmission latency should be less than $20$~ms so that the driver can have enough time to react \cite{7990497}.
Such service requirements will become even more stringent for connected and autonomous vehicles.
For instance, to guarantee the safety of automated overtaking, the communication delay between the overtaking vehicle and surrounding vehicles should be below a few tens of milliseconds as discussed in \cite{5GPPPautomotive} and \cite{8642794}.


To support such stringent service requirements in vehicular networks, there has been a surge of recent works that study the  problem of effective resource allocation for vehicular networks \cite{7289470,8328012,8494751,8341501}.  
For example, the work in \cite{7289470} considers the interference between cellular users and vehicular users and proposes a resource block allocation and power control algorithm to satisfy the latency and reliability requirements of vehicular users.  
Moreover, different from \cite{7289470}, a new resource allocation strategy that additionally considers the latency and reliability requirements under different modulation and coding schemes is proposed in \cite{8328012}.
In addition, the work in \cite{8494751} proposes a proximity and quality-of-service-aware resource allocation framework to minimize the transmission power under reliability and queuing latency constraints for the vehicle-to-vehicle (V2V) communication links.
Furthermore, the authors in \cite{8341501} use extreme value theory to study a power minimization problem subject to both latency and reliability constraints for next-generation vehicular networks. 


While interesting, the body of work in \cite{7289470,8328012,8494751,8341501} does not consider the dependence relationship between delay parameters in a vehicular network.
In fact, for a group of co-existing V2V links, the transmissions will be affected by cross interference and the same set of interferers, leading to a \emph{correlation} between the interference experienced by these links \cite{haenggi2012stochastic}. 
Such correlated interference will lead to a coupling and dependence between the communication delays for these V2V links which can then impact the overall reliability of a vehicular network.
Therefore, instead of considering the wireless performance independently for co-existing V2V links, as done in \cite{7289470,8328012,8494751,8341501}, there is a need to consider the dependence relationship between wireless links when optimizing the vehicular network reliability.

The main contribution of this paper is a novel dependence control mechanism that can improve the overall reliability of a vehicular network by explicitly accounting for the correlation among co-existing V2V links.
In particular, we first characterize the spatial dependence between the delay parameters of concurrent V2V links.
Then, we leverage the concept of a \emph{concordance order} \cite{shaked2007stochastic} from stochastic ordering theory to capture the dependence among delays and further show that a higher concordance can lead to a better overall reliability.
Using such insight, we formulate a power allocation optimization problem and provide an effective solution to maximize the concordance between delay parameters, thereby improving the communication reliability of the vehicular network.
Unlike the work in \cite{6042309} where stochastic orders are solely used to compare the system performance over different channels, we explicitly study the concordance order and propose a power allocation strategy to control the dependence between delays so as to optimize the reliability.  
Furthermore, different from the work in \cite{haenggi2012stochastic} in which the spatio-temporal correlation is captured by a correlation coefficient, here, we measure the dependence based on the concordance order and propose a dependence control mechanism to improve the reliability.    
\emph{To the best of our knowledge, this is the first work that performs resource allocation in wireless networks to control the dependence between the delay parameters of different V2V links so as to optimize the overall reliability of a vehicular network.}
Simulation results validate the theoretical analysis and show that the dependence-based optimization method can yield up to $25$\% gain in terms of the overall reliability compared to a baseline without dependence control.

The rest of this paper is organized as follows. Section \uppercase\expandafter{\romannumeral2} presents the system model. 
In Section \uppercase\expandafter{\romannumeral3}, we introduce the concept of concordance order and use it to solve the reliability optimization problem. Section \uppercase\expandafter{\romannumeral4} provides the simulation results and conclusions are drawn in Section \uppercase\expandafter{\romannumeral5}.

\section{System Model}

Consider a vehicular network that consists of a group $\mathcal{M}$ of $M$ V2V links as well as a set $\mathcal{N}$ of $N$ vehicles that will generate interference to all these V2V links, as shown in Fig. \ref{SystemModel}. 
In this model, the transmitting vehicle in each link sends information (such as speed and location) to the corresponding receiver. 
The receiving vehicle uses the received information to properly adjust its acceleration and driving direction to maintain the safety.

\subsection{Communication Model}
\begin{figure}[!t]
	\centering
	\includegraphics[width=3.2in,height=1.1in]{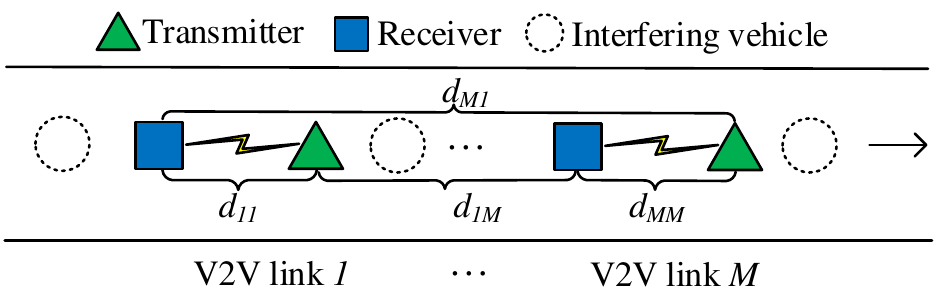}
	\DeclareGraphicsExtensions.
	\vspace{-0.2cm}
	\caption{Highway traffic model that includes two V2V links and a number of interfering vehicles.}
	\label{SystemModel}
	\vspace{-0.4cm}
\end{figure}

To model the location of each vehicle, we use a one-dimensional coordinate system centered on an arbitrarily selected vehicle, as shown in Fig. \ref{SystemModel}.  
In particular, the locations of transmitting and receiving vehicles for any given V2V link $i$ are given, respectively, by $x^t_{i}$ and $x^r_{i}$, $i\in \mathcal{M}$, and the location of interfering vehicle $j\in\mathcal{N}$ is assumed to be $x_{j}$.
Moreover, for the ease of presentation, we let $d_{i,j}=|x^t_{i}-x^r_{j}|$ for $i,j \in \mathcal{M}$.
Therefore, the received power over V2V link $i \in \mathcal{M}$ will be $P_{i}=P^t_{i}g_{i}(d_{i,i})^{-\alpha}$, where $P^t_{i}$ is the transmit power at link $i$, $g_{i}$ is the channel gain, and $\alpha$ is the path loss exponent.
In addition, the interference at the receiver of link $i$ will be given by 
\begin{align}
\label{interference}
I_i =\sum _{j\in \mathcal{M} \setminus i}P^t_{j}g_{j,i}(d_{j,i})^{-\alpha}+ \sum_{k\in\mathcal{N}}P^{t}_{k}g_{k,i}|x_{k}-x_{i}^r|^{-\alpha},
\end{align}
where $g_{k,i}$ is the channel gain between interfering vehicle $k$ and the receiving vehicle of link $i$.
Note that the first term on the right-hand side of (\ref{interference}) captures the cross-interference within the set $\mathcal{M}$ of V2V links, and the second term represents the interference from the common set $\mathcal{N}$ of interfering vehicles.
Without loss of generality, we assume the transmission power of interfering vehicles to be equal, i.e., $P^{t}_{k} = P_{c}, \forall k \in \mathcal{N}$.
Accordingly, we can derive the rate of V2V link $i$ as
\begin{align}
\label{throughtput}
R_{i}= \omega \log_{2} \left( 1 + \frac{P_{i}}{\omega N_{0}+I_{i}}\right),
\end{align} 
where $\omega$ is the bandwidth of the wireless channel, and $N_{0}$ is the noise power spectral density. 
Using (\ref{throughtput}), we can obtain the corresponding transmission delay of link $i$ as $t_{i} = S /R_{i}$, where $S$ is the size of the transmitted packet in bits. 

\subsection{Problem Formulation}
We define $\tau_i$ as the target delay for any V2V link $i \in \mathcal{M}$. Then, we have two approaches for defining the reliability performance of the vehicular network. 
One approach is to directly compare the transmission delay  $t_{i}$ with the threshold $\tau_{i}, i \in \mathcal{M}$.
The other approach is to derive $\mathbb{P}(t_{i} \leq \tau_{i})$, which is the probability that $t_{i}$ is smaller than $\tau_{i}$.
The second approach has been very popular in the literature \cite{7289470,8328012,8494751,8341501} where it is used as a constraint in the resource allocation problem.


However, the aforementioned two approaches ignore the dependence between the delay parameters of the V2V links which stems from: a) Interference from the same group of interfering vehicles in the set $\mathcal{N}$, and b) Cross-interference between V2V links in the set $\mathcal{M}$. 
Therefore, quantifying the system's reliability via these two aforementioned, conventional approaches may not capture the actual reliability of real-world vehicular networks. 
To consider the dependence between all delay components $t_i, \forall i \in \mathcal{M}$, we seek to optimize the joint reliability, i.e., $\mathbb{P}(t_{1}\leq \tau_{1},..., t_{M} \leq \tau_{M})$, as follows:
\begin{align} \label{Opt1}
&\max_{P^t_{1},...,P^t_{M}}  \mathbb{P}(t_{1}\leq \tau_{1},...,t_{M} \leq \tau_{M})   \\
&\hspace{0.13in}\text{s.t.} \hspace{0.04in}  0 \leq P^t_{i} \leq P^{\text{max}},\hspace{0.08in} i \in \mathcal{M}, \label{OptCon1} 
\end{align}
where constraint (\ref{OptCon1}) ensures that the transmit power will not exceed a maximum threshold.  

To solve the optimization problem in (\ref{Opt1}), we need to find the joint reliability $\mathbb{P}(t_{1}\leq \tau_{1},...,t_{M} \leq \tau_{M})$.
However, deriving the joint reliability expression is challenging, since it  depends on the spatial distribution of interfering vehicles in $\mathcal{N}$ as well as on the fading channels for all $M$ V2V links.
Moreover, even if the spatial distribution of interfering vehicles and the fading channels are specified, it is still difficult to find a tractable  closed-form expression for joint reliability due to the complex channel power distribution.

In the following section, we propose a new dependence control framework that can be used to solve the joint reliability optimization problem in (\ref{Opt1}).
In particular, we first explicitly link the dependence between the delay parameters with the joint reliability of the vehicular network.
Then, based on this analysis, we reformulate the optimization problem and use tools from convex optimization to find an efficient solution to the reformulated problem. 

\section{Dependence Control based Reliability Optimization}
In this section, we first introduce the concept of concordance order from stochastic ordering theory and apply it to our reliability optimization problem. 
Then, we show that objective function (\ref{Opt1}) is directly linked to the concordance between the delay parameters of the V2V communication links.  
We later use the concordance measure tool to reformulate the optimization problem in (\ref{Opt1}) and adapt the dual update method to solve the new problem. 
\subsection{Concordance Order: Preliminaries}
In probability theory and statistics, stochastic ordering theory provides methods to compare random variables or random vectors \cite{shaked2007stochastic}. 
For example, for two random variables $X$ and $Y$, $X$ is said to be smaller than $Y$ in the stochastic order when $\mathbb{E}f(X) \leq_{\text{st}} \mathbb{E}f(Y)$ holds for any increasing function $f$ and the expectations $\mathbb{E}f(X)$ and $\mathbb{E}f(Y)$ exist \cite{muller2002comparison}.
By modifying the properties of function $f$, we can further classify the stochastic orders into convex orders, supermodular orders, and others. 
Among these extensions, one of the most popular ordering methods is the so-called \emph{concordance order} that ranks random vectors based on the dependence structure between them.
In particular, random variables in a vector are in concordance if ``large'' values of one coincide with ``large'' values of the other and ``small'' values of one with ``small'' values of the other.
Moreover, if one random vector $\boldsymbol{X}=(X_{1},...,X_{M})$ is said to be smaller than another vector $\boldsymbol{Y}=(Y_{1},...,Y_{M})$ in concordance order (written as $\boldsymbol{X} \leq_{c} \boldsymbol{Y}$), we will have  
\begin{align}
\label{concordance}
\mathbb{P}(X_1 \!\leq\! s_{1},...,X_{M} \!\leq\! s_{M}) \leq \mathbb{P}(Y_{1} \!\leq\! s_{1},...,Y_M \!\leq\! s_{M}),
\end{align}
for $s_i \in \{-\infty,\infty\}, \forall i\in \mathcal{M}$ \cite{muller2002comparison}.  

For our problem in (\ref{Opt1}) and (\ref{OptCon1}), we seek to maximize the joint reliability $\mathbb{P}(t_{1}\leq \tau_{1},...,t_{M} \leq \tau_{M})$ by optimizing the power allocation of V2V links. 
Such optimization problem aims to find a power allocation scheme such that the associated delay vector $(t'_{1},...,t'_{M})$ for all $M$ links can meet the following condition:   
\begin{align}
\mathbb{P}(t_{1}\leq \tau_{a},...,t_{M}\leq \tau_{M}) \leq \mathbb{P}(t'_{1} \leq \tau_{1},...,t'_M \leq \tau_{M}),
\end{align} 
where $(t_{1},...,t_{M})$ is the associated delay vector for any other transmission power scheme (within the feasibility set) allocated to $M$ V2V links.
According to the definition of a concordance order in (\ref{concordance}), solving the optimization problem in (\ref{Opt1}) and (\ref{OptCon1}) is thereby equivalent to finding a power allocation strategy that can maximize the concordance order of the delay vector $(t_{1},...,t_{M})$.
In the following subsection, we introduce a method based on the theory of copula to measure the concordance in the delay vector.

\subsection{Copula-based Concordance Measure}
To measure the concordance between delay parameters $t_{i}$, $i \in \mathcal{M}$, we use \emph{copula theory}. 
This is because copulas are capable of measuring any non-linear and time varying dependence between delay variables \cite{nelsen2007introduction}, introduced by shadowing, Doppler shift, and mobility which are factors that are inherent to vehicular communications.  
In particular, for a random vector, a copula captures the dependence structure between marginal parameters.
Moreover, a copula can be expressed as a function which links a joint distribution function with its one-dimensional marginal distribution functions \cite{nelsen2007introduction}. 
Mathematically, given a multivariate joint distribution function $H$ with marginal cumulative distribution function $F_{i}, i \in \mathcal{M}$, the $M$-dimensional copula is expressed as $C(u_1,...,u_M)= H(F_{1}^{-1}(u_{1}),...,F_{M}^{-1}(u_{M}))$, where $(u_1,...,u_{M})$ are restricted in $[0,1]^M$ and $u_i$ follows a uniform distribution. 
Another type of commonly used copulas is called the survival copula, mathematically defined as $\hat{C}(u_1,...,u_M)=H(1-F_{1}^{-1}(u_{1}),...,1-F_{M}^{-1}(u_{M}))$.

Moreover, copula theory provides a foundation to measure the concordance between marginal parameters in a random vector. 
In particular, there are many copula-based methods to measure the concordance between random variables, such as Kendall's $\tau$ function, Spearman's $\rho$ function, Gini's $g$ function, and Blomqvist's $\beta$ function \cite{nelsen2007introduction}.
These measurement methods are all linked to each other, and, here for a total of $M$ V2V links, we use the multivariate version of Blomqvist's $\beta$ function, given by \cite{ubeda2005multivariate}
\begin{align}
\label{multivariatebeta}
\beta = \frac{2^{M-1}[C(1/2,...,1/2)+\hat{C}(1/2,...,1/2)]-1}{2^{M-1}-1}.
\end{align} 
Note that, although $\beta$ depends on the value of copula at the center of $[0,1]^M$, as shown in (\ref{multivariatebeta}), it can provide an accurate approximation of other copula-based measurements \cite{nelsen2007introduction}. 
In the following theorem, we use stochastic geometry to derive the explicit expression of Blomqvist's $\beta$ function between the delay parameters in our vehicular network.

\begin{theorem}
	\label{theorem1} Given that the distribution of interfering vehicles follows a one-dimensional Poisson point process (1-D PPP) with density $\lambda$ and the channels of V2V links and interfering links are assumed to be independent Rayleigh fading channels, the Blomqvist's $\beta$ function (concordance measurement) of delay parameters is given by:
	\begin{align}
		&\beta(P_1^t,...,P_M^t) = \nonumber \\ 
		&\frac{2^{M\!-\!1}\![H\!\left(\!F^{-1}_{1}\!\!\left(\frac{1}{2}\right)\!,\!...,\!F^{-1}_{M}\!\!\left(\frac{1}{2}\right)\!\right)\!\!+\!\!H\!\left(\!1\!\!-\!\!F^{-1}_{1}\!\!\left(\frac{1}{2}\right)\!,\!...,\!1\!\!-\!\!F^{-1}_{M}\!\!\left(\frac{1}{2}\right)\!\right)]\!\!-\!\!1}{2^{M-1}-1}, 
	\end{align}
	where 
	\begin{align}
	&F_{i}(v) =\prod_{j\in\mathcal{M}\setminus i} \left(\frac{1}{1+\frac{(2 ^{\frac{S}{\omega v}}\!-\!1) P^t_{j}(d_{j,i})^{-\alpha}}{P^t_{i}(d_{i,i})^{-\alpha}}}\right) \times \nonumber \\ &\hspace{0.1in} \exp\left(-\lambda \int_{-\infty}^{\infty}1- \frac{1}{1+\frac{ (2 ^{\frac{S}{\omega v}}\!-\!1)P_{c} |x_{k}\!-\!x_{i}^r|^{-\alpha} }{P^t_{i}(d_{i,i})^{-\alpha}}}dx_{k}  \right),  
	\end{align}
	and 
	{\small\begin{align}
	&H(u_1,...,u_M) = \nonumber \\ 
	&\!\!\!\!\prod_{j\in\mathcal{M}\setminus 1}\!\!\frac{1}{1\!\!+\!\!\frac{ (2 ^{\frac{S}{\omega u_1}}\!-\!1) P^t_{j}(d_{j,1})^{-\alpha}}{P^t_{1}(d_{1,1})^{-\alpha}}} \!\times\!\!...\!\! \times\!\! \!\!\!
	\prod_{j\in\mathcal{M}\setminus M} \!\!\frac{1}{1\!\!+\!\!\frac{ (2 ^{\frac{S}{\omega u_M}}\!-\!1) P^t_{j}(d_{j,M})^{-\alpha}}{P^t_{M}(d_{M,M})^{-\alpha}}}\! \times\! \nonumber \\ 
	&\!\!\exp\!\!\left( \!\!-\!\lambda\!\! \int_{-\infty}^{\infty}\!\!\!\!\!\!\!\!\! 1\!\!-\!\! \frac{dx_{k}}{\left(\!\!1\!\!+\!\!\frac{ (2 ^{\!\frac{S}{\omega u_1}\!}\!-\!1)\! P_{c}\!|\!x_{k}\!-\!x^{r}_{1}\!|^{\!-\!\alpha}}{P^t_{1}(d_{1,1})^{-\alpha}}\!\!\right)\!\!\!\times\!\!...\!\!\times\!\!\left(\!\!1\!\!+\!\!\frac{ (2 ^{\!\frac{S}{\omega u_M}\!}\!-\!1) \!P_{c}\!|\!x_{k}\!-\!x^{r}_{M}\!|^{\!-\!\alpha}}{P^t_{M}(d_{M,M})^{-\alpha}}\!\!\right)} \!\!\! \right).
	\end{align}}
	\begin{proof}
		Please refer to Appendix \ref{appendix1}.
	\end{proof}
\end{theorem}

As the Blomqvist's $\beta$ function is directly linked to the concordance between $t_{i}, i\in \mathcal{M}$, the theoretical expression of Blomqvist's $\beta$ function in Theorem \ref{theorem1} allows us to analyze the concordance change in terms of the transmission power.
Based on the relationship between concordance and joint reliability, we can also determine the transmission power of V2V links so that the joint reliability is maximized. 
Thus, in the following subsection, we reformulate the proposed optimization problem by replacing the objective function in (\ref{Opt1}) with the concordance measure, i.e., Blomqvist's $\beta$ function. 
We also provide an effective approach to solving the reformulated problem.  

\subsection{Problem Reformulation and Proposed Solution}
According to \cite{nelsen2007introduction}, the Blomqvist's $\beta$ function is an increasing function in terms of the concordance between delay parameters $t_{i}, i\in \mathcal{M}$. 
Hence, we can perform transmission power allocation for V2V links in order to maximize the concordance, thereby optimizing the joint probability $\mathbb{P}(t_{1}\leq \tau_{1},...,t_{M} \leq \tau_{M})$. 
In other words, the original problem in (\ref{Opt1}) and (\ref{OptCon1}) is equivalent to the following problem: 
\begin{align} \label{Opt2}
&\max_{P^t_{1},...,P^t_{M}} \beta(P_1^t,...,P_M^t)  \\
&\hspace{0.13in}\text{s.t.} \hspace{0.04in}  0 \leq P^t_{i} \leq P^{\text{max}},\hspace{0.08in} i \in \mathcal{M}. \label{conre1} 
\end{align}

As the convexity of the reformulated optimization problem is challenging to  determine, we use the dual updated method \cite{1658226} to obtain an efficient sub-optimal solution. 
In particular, we iteratively update the Lagrange multipliers in the Lagrange function so as to find the optimal values for Lagrange multipliers. 
Then, given the optimal values of the Lagrange multipliers, we calculate the sub-optimal values of transmission power for $M$ V2V links in the system by solving the dual optimal problem.  
First, we obtain the Lagrange function as 
\begin{align}
\label{Lagrange}
L(\theta_{1},&...,\theta_{M},\vartheta_1,...,\vartheta_M)= \nonumber \\ &\beta(P_1^t,...,P_M^t)\!+\sum_{i=1}^{M} \theta_{i}P_i^t \!+\!
\sum_{j=1}^{M}\vartheta_j (P^{\text{max}}\!\!-\!\!P_j^t), 
\end{align}
where $\theta_{i}$ and $\vartheta_j, i,j\in \mathcal{M},$ are the Lagrange multipliers for constraint (\ref{conre1}). 
Note that although the dual problem in (\ref{Lagrange}) is always convex in terms of $\theta_{i}$ and $\vartheta_j, i,j\in \mathcal{M}$, $L(\theta_{1},...,\theta_{M},\vartheta_1,...,\vartheta_M)$ is not differentiable. 
Thus, we derive the subgradient of $L(\theta_{1},...,\theta_{M},\vartheta_1,...,\vartheta_M)$ as follows: 
\begin{align} \label{sub1}
&\Delta \theta_{i} = (P^t_{i})^*, \Delta \vartheta_i = P_i^{\text{max}} - (P^t_{i})^*, 
\end{align}
where $(P^t_{i})^*$ is the optimal solution to $P_i^{t}, i \in \mathcal{M}$.
The proof of subgradients is similar to the one provided in \cite{1658226} and is omitted here. 
With the updated Lagrange multipliers, we can update the values of $P^t_{i}, i\in \mathcal{M}$, by solving the dual optimization problem, following the framework in \cite{1658226}. Moreover, we choose the ellipsoid method to find the dual variables, and all variables will converge in $\mathcal{O}(49\log(1/\eta))$ iterations where $\eta$ is the accuracy \cite{boyd2004convex}.

Since the reformulated concordance optimization problem is equivalent to the  problem defined by (\ref{Opt1}) and (\ref{OptCon1}), the base station or roadside unit can use the sub-optimal solution obtained from the dual update method to allocate transmission power for V2V links so as to improve the joint reliability performance of the vehicular network.
\vspace{-0.05in}
\section{Simulation Results and Analysis}\vspace{-0.05in}
\begin{table}[!t]
	\large
	\begin{center}
		\caption{\small Simulation parameters.}
		\vspace{-0.6cm}
		\label{table_example}
		\resizebox{9cm}{!}{
			\begin{tabular}{|c|c|c|}
				\hline
				\textbf{Parameter} & \textbf{Description} & \textbf{Value} \\  \hline
				$d_{a,a},d_{b,b}$ & Distance within each V2V link & $5$, $5$~m \\ \hline 
				{\shortstack{$d_{a,b},d_{b,a}$\vspace{0.05in}}} & {\shortstack{\\Distance between the receiver in one link \\ and the transmitter in another link}} & {\shortstack{$5$, $15$~m\vspace{0.05in}}} \\\hline
			    $P^{\text{max}}$ & Maximum transmission power & $27$~dBm \cite{zeng2018joint}\\ \hline  
				$P_{c}$ & Transmission power of interfering vehicles & $25$~dBm \\ \hline 
				$\alpha$  & Path loss exponent & $3$ \\ \hline
				$N_0$ & Noise power spectral density  & $-174$~dBm/Hz  \\ \hline
				$S$ & Packet size & $3,200$~bits \\ \hline 
				$\omega$ & Bandwidth & $20$~MHz \\ \hline 
				$\tau_{1},\tau_{2}$ & Delay requirements & $13.9,13.9$~ms  \cite{zeng2018joint}\\ \hline
		\end{tabular}}
		\vspace{-0.6cm}
	\end{center}
\end{table}

For our simulations, we consider a vehicular network with two V2V links, i.e., $M = 2$, as well as a group of interfering vehicles where their spatial distribution follows a 1D-PPP. 
In particular, we first numerically validate the Blomqvist's $\beta$ function derived in Theorem \ref{theorem1}. 
Moreover, we study the performance gain achieved by the V2V network when performing dependence control between the delay parameters of V2V links.
Without loss of generality, we assume that the distance between the transmitter and receiver in these two V2V links is equal, i.e., $d_{1,1}=d_{2,2}$.
All simulation parameters are summarized in Table \ref{table_example}.
We also consider a road segment of $20$~km. All statistical results are averaged over a large number of independent runs.

\begin{figure}[!t]
	\centering
	\includegraphics[width=2.8in,height=2.1in]{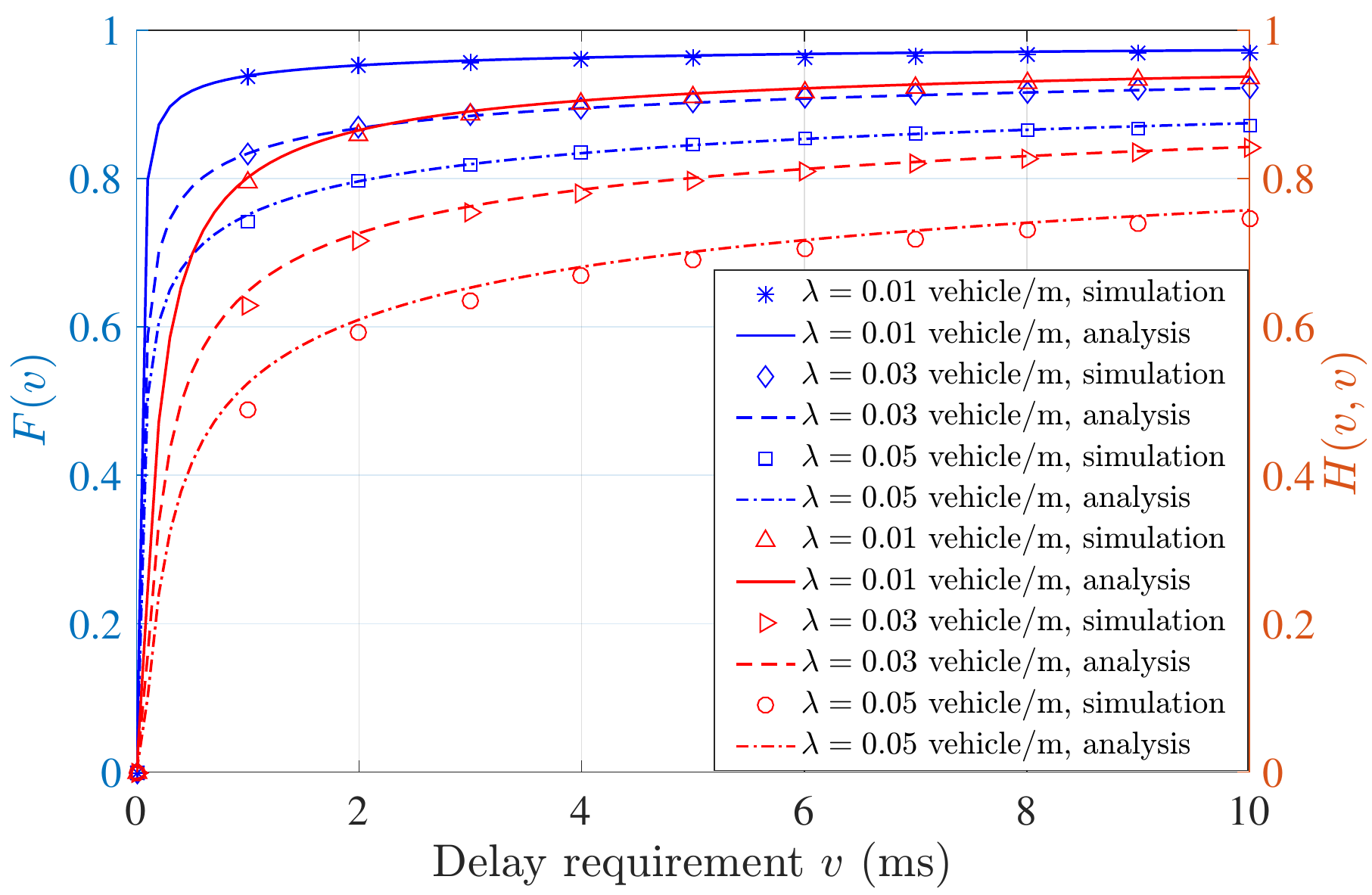}
	\DeclareGraphicsExtensions.
	\vspace{-0.3cm}
	\caption{Validation of $F$ and $H$ derived in Theorem \ref{theorem1}.}
	\label{validate}
	\vspace{-0.4cm}
\end{figure}

\begin{figure}[!t]
	\centering
	\includegraphics[width=2.6in,height=2.0in]{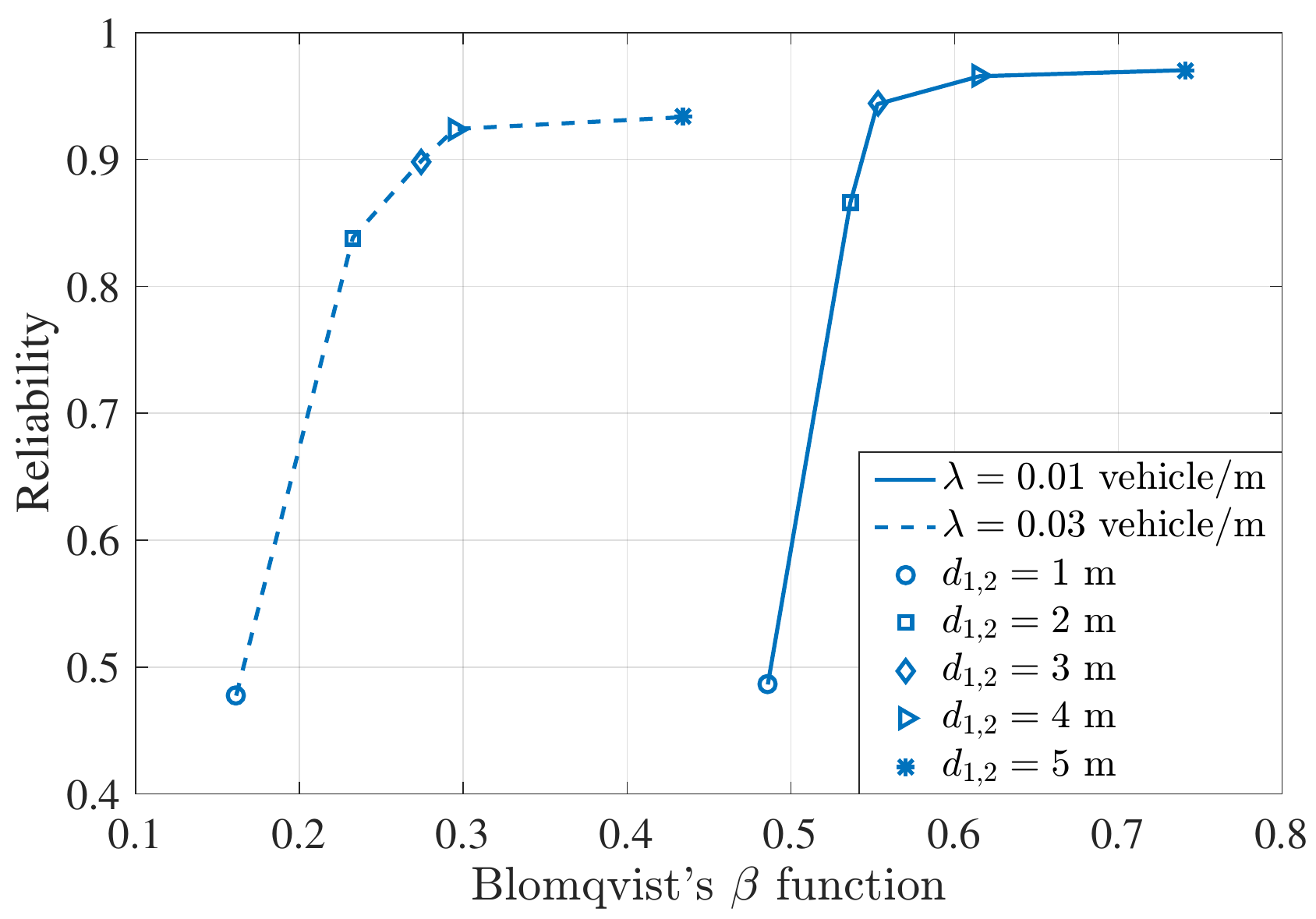}
	\DeclareGraphicsExtensions.
	\vspace{-0.4cm}
	\caption{Reliability performance versus Blomqvist's $\beta$ function for vehicular networks with $\lambda=0.01$~vehicle/m and $\lambda=0.03$~vehicle/m.}
	\label{beta}
	\vspace{-0.4cm}
\end{figure}

Fig. \ref{validate} shows the distribution of functions $F$ and $H$ derived in Theorem \ref{theorem1} for V2V networks with different distribution density of interfering vehicles. 
In particular, according to the empirical data collected by the California PATH Program, we choose the density of interfering vehicles within a range from $0.01$~vehicle/m to $0.05$~vehicle/m \cite{BerkeleyHighwayLabWebsite}. 
As observed in Fig. \ref{validate}, the simulation results match the derived functions $F$ and $H$ in Theorem \ref{theorem1}, guaranteeing the effectiveness of calculating the concordance measurement, i.e., Blomqvist's $\beta$ function.
Moreover, Fig. \ref{validate} shows that, for the same network setting, $F(v)$ always outperforms $H(v,v)$. 
This is due to the fact that $F(v)$ only captures the probability of the communication delay of one link being below than the delay threshold $v$, i.e., $\mathbb{P}(t_i \!\leq\! v), i \!\in\! \{1,2\}$.
However, $H(v,v)$ represents the probability that the delays of links $1$ and $2$ are both smaller than the delay threshold $v$, i.e., $\mathbb{P}(t_1 \!\leq\!  v,\hspace{0.02in} t_2 \leq v)$.
According to basic probability theory, $\mathbb{P}(t_i \!\leq\! v) \geq \mathbb{P}(t_1 \! \leq\! v,\hspace{0.02in} t_2 \!\leq\! v), i \!\in\! \{1,2\},$  always exist.

Fig. \ref{beta} shows the reliability of two networks with $\lambda=0.01$~vehicle/m and $\lambda=0.03$~vehicle/m when the concordance, i.e., the value of Blomqvist's $\beta$ function, increases. 
In particular, we keep increasing the value of the distance $d_{1,2}$ from $1$~m to $5$~m so as to calculate the corresponding values of the reliability and Blomqvist's $\beta$ function.
As observed from Fig. \ref{beta}, when Blomqvist's $\beta$ increases, the reliability of the vehicular network also increases, verifying the intuition that a high concordance leads to a higher reliability.  
Moreover, Fig. \ref{beta} shows that, when choosing the same value of $d_{1,2}$, the reliability is always better for the network with a smaller density $\lambda$.
For example, when choosing $d_{1,2}=5$~m, the reliability is approximately $97$\% for the network with $\lambda=0.01$~vehicle/m, whereas the counterpart for the system with $\lambda=0.03$~vehicle/m is around $92$\%.
This is due to the fact that, with more interfering vehicles in the network, interference increases and then results in a lower reliability. 
\begin{figure}[!t]
	\centering
	\includegraphics[width=2.8in,height=2.1in]{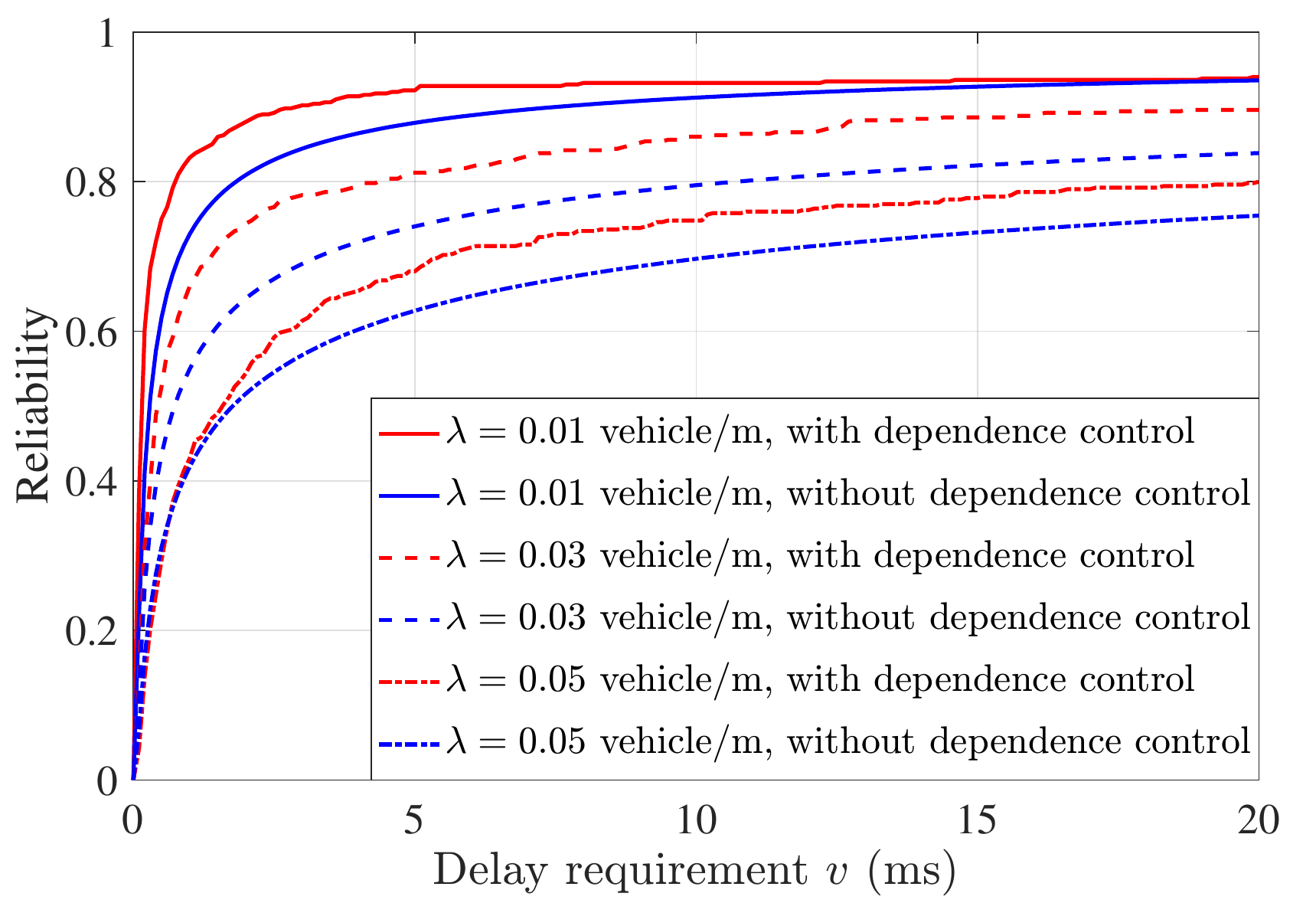}
	\DeclareGraphicsExtensions.
	\vspace{-0.4cm}
	\caption{Reliability performance for networks with dependence control and without dependence control.}
	\label{Optimize}
	\vspace{-0.4cm}
\end{figure}
	

Fig. \ref{Optimize} compares the reliability performance of V2V networks with and without dependence control. 
In particular, we choose the random power allocation scheme as a baseline without dependence control. 
As shown in Fig. \ref{Optimize}, for the same network setting, the reliability performance for the networks with dependence control is always better than counterparts without dependence control. 
In particular, when $\lambda=0.03$~vehicle/m and delay requirements $\tau_1 = \tau_2 =13.9$~ms \cite{zeng2018joint}, the reliability of the network with dependence control is around $89$\% whereas the reliability of the network without dependence control is approximately $81$\%, which is a significant gain in terms of the overall reliability.
Moreover, when $\lambda=0.03$~vehicle/m and the delay requirement is set to $v=1$~ms, the network with dependence control can achieve up to $25$\% reliability gain compared with the network without dependence control. 
In addition, as shown in Fig. \ref{Optimize}, the reliability gain achieved by performing dependence control will first increase and then decrease when the traffic density $\lambda$ increases. 
This stems from the fact that, when the traffic density $\lambda$ is low, the interference will be limited and the delay requirement can be easily met for networks with and without dependence control. 
When $\lambda$ increases and consequently interference increases, the delay requirement becomes more stringent for the network without dependence control. 
However, networks with dependence control can optimize the power allocation to maximize the overall reliability performance.
Furthermore, Fig. 4 shows that when the traffic is too dense, the reliability gain yielded by the dependence control naturally decreases due to the severe interference generated from other interfering vehicles.

\vspace{-0.05in}
\section{Conclusion}\vspace{-0.05in}
In this paper, we have proposed a novel power allocation framework based on the dependence control between delay parameters for V2V links to optimize the network reliability. 
We have used the notion of concordance from stochastic ordering theory to obtain insights on the relationship between the dependence and the overall reliability. 
Based on the dependence structure modeled by stochastic geometry, we have proposed a new optimization problem to maximize the concordance and optimize the reliability of the vehicular network. 
A dual update method has been adopted to obtain a sub-optimal solution of the power allocation for V2V links. 
Simulation results have corroborated the theoretical analysis and shown the performance gain resulting from the proposed dependence control scheme for the vehicular network.      

\vspace{-0.1in}
\appendix\vspace{-0.05in}
\subsection{Proof of Theorem \ref{theorem1}}
\label{appendix1}
The marginal cumulative distribution function of the delay experienced by V2V link $ i \in \mathcal{M}$ can be derived as follows \vspace{-0.05in} 
{\small\begin{align}
	\label{13}
&\!\!F_{i}(v) \stackrel{(a)}{\approx} \mathbb{P}\left(\frac{S}{R_{i}} \leq v\right)
\nonumber \\
&\!\!\!=\!\!\mathbb{P}\!\left(\!\frac{P^t_{i}g_{i}(d_{i,i})^{-\alpha}}{\sum _{j\in \mathcal{M} \setminus i}\!P^t_{j}g_{j,i}(d_{j,i})^{-\alpha}\!+\! \sum_{k\in\mathcal{N}}\!P_{c}g_{k,i}|x_{k}\!\!-\!\!x_{i}^r|^{-\alpha}}\! \geq\! 2 ^{\frac{S}{\omega v}}\!\!-\!\!1 \right)
\nonumber \\ 
&\!\!\!=\!\! \mathbb{P}\!\!\left(\!\!g_i\!\! \geq \!\!\frac{\!(2 ^{\frac{S}{\omega v}}\!\!-\!\!1)(\sum _{j\in \mathcal{M} \setminus i}\!P^t_{j}g_{j,i}(\!d_{j,i}\!)^{\!-\!\alpha}\!\!+\!\! \sum_{k\in\mathcal{N}}\!P_{c}g_{k,i}|x_{k}\!\!-\!\!x_{i}^r|^{\!-\!\alpha})}{P^t_{i}(d_{i,i})^{-\!\alpha}}\!\right)
\nonumber \\ 
&\!\!\!\stackrel{(\!b\!)}{=} \!\! \mathbb{E} \!\exp\!\!\left(\!\!\frac{\!-(2 ^{\!\frac{S}{\omega v}\!}\!\!-\!\!1)(\sum _{j\in \!\mathcal{M} \setminus i}\!P^t_{j}g_{j,i}(\!d_{j,i}\!)^{\!-\!\alpha}\!\!\!+\!\! \sum_{k\in\mathcal{N}}\!\!P_{c}g_{k,i}|x_{k}\!\!-\!\!x_{i}^r|^{\!-\!\alpha})}{P^t_{i}(d_{i,i})^{\!-\!\alpha}} \!\! \right)  
\nonumber \\ 
&\!\!\!=\! \mathbb{E}_{\mathcal{N}}\Bigg[ \prod_{j\in\mathcal{M}\setminus i}\exp\left(-\frac{g_{j,i} (2 ^{\frac{S}{\omega v}}\!-\!1)\! P^t_{j}(d_{j,i})^{-\alpha}}{P^t_{i}(d_{i,i})^{-\!\alpha}} \right) \times \nonumber \\ &\hspace{0.45in}\prod_{k\in\mathcal{N}}\exp\left( -\frac{g_{k,i} (2 ^{\frac{S}{\omega v}}\!-\!1)P_{c} |x_{k}\!-\!x_{i}^r|^{-\!\alpha} }{P^t_{i}(d_{i,i})^{-\alpha}}\!\right)\Bigg] \nonumber \\ 
&\!\!\!\stackrel{(c)}{=} \!\!\!\prod_{j\in\mathcal{M}\setminus i}\!\!\left(\!\!\frac{1}{1\!+\!\frac{ (2 ^{\frac{S}{\omega v}}\!-\!1) P^t_{j}(d_{j,i})^{-\alpha}}{P^t_{i}(d_{i,i})^{-\alpha}}}\!\!\right)\mathbb{E}_{\mathcal{N}} \!\!\prod_{k\in\mathcal{N}}\!\!\left(\!\!\frac{1}{1\!+\!\frac{ (2 ^{\frac{S}{\omega v}}\!-\!1)P_{c} |x_{k}\!-\!x_{i}^r|^{\!-\alpha} }{P^t_{i}(d_{i,i})^{-\alpha}}}\!\!\right) \nonumber \\ 
&\!\!\!\stackrel{(d)}{=} \prod_{j\in\mathcal{M}\setminus i} \left(\frac{1}{1+\frac{(2 ^{\frac{S}{\omega v}}\!-\!1) P^t_{j}(d_{j,i})^{-\alpha}}{P^t_{i}(d_{i,i})^{-\alpha}}}\right) \times \nonumber \\ 
&\!\!\!\hspace{0.2in} \exp\left(-\lambda \int_{-\infty}^{\infty}1- \frac{1}{1+\frac{ (2 ^{\frac{S}{\omega v}}\!-\!1)P_{c} |x_{k}\!-\!x_{i}^r|^{-\alpha} }{P^t_{i}(d_{i,i})^{-\alpha}}}dx_{k}  \right),
\vspace{-0.1in}\end{align}}where in (a), we assume that the signal-to-interference-plus-noise-ratio of link $i$ can be approximated as its signal-to-interference-ratio.
The changes in (b) and (c) follow the assumption of independent Rayleigh fading channels, while (d) can be explained by the probability generating functional (PGFL) of a PPP \cite{haenggi2012stochastic}. 

The joint distribution for the delay parameters can be derived similarly, but we need to take into account the dependence between them. The proof is as follows\vspace{-0.05in}
{\small\begin{align}
&H(u_1,...,u_M)\approx \mathbb{P} \left( \frac{S}{R_{1}} \leq u_{1},..., \frac{S}{R_{M}} \leq u_M   \right) \nonumber \\ 
&=\!\!\!\!\prod_{j\in\mathcal{M}\setminus 1}\!\!\frac{1}{1\!\!+\!\!\frac{ (2 ^{\frac{S}{\omega u_1}}\!-\!1) P^t_{j}(d_{j,1})^{-\alpha}}{P^t_{1}(d_{1,1})^{-\alpha}}} \!\times\!\!...\!\! \times\!\! \!\!\!\!
\prod_{j\in\mathcal{M}\setminus M} \!\!\frac{1}{1\!\!+\!\!\frac{ (2 ^{\frac{S}{\omega u_M}}\!-\!1) P^t_{j}(d_{j,M})^{-\alpha}}{P^t_{M}(d_{M,M})^{-\alpha}}}\! \times\! \nonumber \\ 
&\mathbb{E} \prod_{k\in\mathcal{N}}\left(\frac{1}{1\!\!+\!\!\frac{ (2 ^{\frac{S}{\omega u_1}}\!-\!1) P_{c}|x_{k}\!-\!x^{r}_{1}|^{\!-\!\alpha}}{P^t_{1}(d_{1,1})^{\!-\!\alpha}}} \!\times...\times\! \frac{1}{1\!\!+\!\!\frac{ (2 ^{\frac{S}{\omega u_M}}\!-\!1) P_{c}|x_{k}\!-\!x^{r}_{M}|^{\!-\!\alpha}}{P^t_{M}(d_{M,M})^{\!-\!\alpha}}}  \right) \nonumber \\ 
&\stackrel{(a)}{=}\!\!\!\!\prod_{j\in\mathcal{M}\setminus 1}\!\!\frac{1}{1\!\!+\!\!\frac{ (2 ^{\frac{S}{\omega u_1}}\!-\!1) P^t_{j}(d_{j,1})^{\!-\!\alpha}}{P^t_{1}(d_{1,1})^{\!-\!\alpha}}} \times\!\!...\!\! \times\!\! \!\!\!\!
\prod_{j\in\mathcal{M}\setminus M} \!\!\frac{1}{1\!\!+\!\!\frac{ (2 ^{\frac{S}{\omega u_M}}\!-\!1) P^t_{j}(d_{j,M})^{\!-\!\alpha}}{P^t_{M}(d_{M,M})^{\!-\!\alpha}}}\! \times\! \nonumber \\ 
&\exp\!\left( \!\!-\!\lambda\!\! \int_{-\infty}^{\infty}\!\!\!\!\!\!\!\!\! 1\!\!-\!\! \frac{dx_{k}}{\left(\!\!1\!\!+\!\!\frac{ (2 ^{\!\frac{S}{\omega u_1}\!}\!-\!1)\! P_{c}\!|\!x_{k}\!-\!x^{r}_{1}\!|^{\!-\!\alpha}}{P^t_{1}(d_{1,1})^{-\alpha}}\!\!\right)\!\!\!\times\!\!...\!\!\times\!\!\left(\!\!1\!\!+\!\!\frac{ (2 ^{\!\frac{S}{\omega u_M}\!}\!-\!1) \!P_{c}\!|\!x_{k}\!-\!x^{r}_{M}\!|^{\!-\!\alpha}}{P^t_{M}(d_{M,M})^{-\alpha}}\!\!\right)} \!\!\! \right), 
\end{align}}where we have omitted intermediate steps which are analogous to (\ref{13}). The change in (a) is based on the PGFL of the PPP.
According to the definition of Blomqvist’s $\beta$ function in (\ref{multivariatebeta}), we can derive the results in Theorem \ref{theorem1}.\vspace{-0.05in} 
\def\baselinestretch{0.65}
\bibliographystyle{IEEEtran}

\end{document}